\documentstyle[12pt]{article}
\begin{document}

 \title{THE CASIMIR EFFECT AS A POSSIBLE SOURCE OF COSMIC ENERGY}
 \author{Igor Yu.Sokolov \\
 Dept. of  Physics, Toronto University, Ontario, M5S 1A7  Canada}
 \date{}
 \maketitle

 \begin{abstract}
 Energy production due to the  Casimir effect is considered
 for the case of a superdense state of matter,
 which can appear in such cosmological objects as white dwarfs, neutron
 stars, quasars and so on. The energy
 output produced  by the Casimir effect during the creation of a
 neutron star turns out to be sufficient to explain nova and
 supernova explosions. It is shown that the Casimir effect might
 be a possible source of the huge energy output of quasars.
 \end{abstract}

 \vfill{} \hskip14cm {\bf UTPT-94-12}

 \newpage

 The aim of the present work is to attract the attention of physicists
 to another possible source of energy in the cosmic scale, namely the
 Casimir effect. This effect is well-known in physics (see,  e.g., [1-3]).
 It arises from the energy shift of zero-point vacuum fluctuations
 of an electromagnetic (e.m.) field due to non-trivial boundary conditions
 for the field. Usually the so-called {\it surface} Casimir effect
 is considered,
 which takes place when the boundary condition is given on a definite
 surface. An example of the surface Casimir effect is the appearance
 of a small attractive force between two uncharged metal plates.
 At the same time, in his work [1], Casimir mentioned a {\it volume}
 effect. Consider the latter in more detail. Let there be zero boundary
 conditions
 for an electromagnetic field in a volume $V$ (taking the volume $V$
 to be filled by an ideal conductor). Calculate now the shift
 of energy density of the field for the volume $V$
 relative to the case of the absence of the volume. It is equal to the mean
 value of the 00-component of the stress-energy tensor of the
 quantum electromagnetic
 field inside the volume $V$

 \begin{equation}
 \Delta \epsilon_{vac} =<0| T_{00} |0>_{inside \ of\  the \ conductor}
 - <0| T_{00} |0>_{empty \ space} = -\hbar c \int \omega
 \frac{d^3 {\bf k}}{(2\pi)^3} \; ,
 \end{equation}

 \noindent where $\omega =|{\bf k}|$ and $ {\bf k}$ is the wave vector of the
 electromagnetic field (dimension of ${\bf k}$ is 1/length).

 One may see that the integral (1) is equal to infinity. So, the reasonable
 final result should depend on a dimensional cut-off parameter. It is
 the reason why  the surface effect is usually considered, where the
 final result does not depend, as a rule, on any cut-offs and where
 divergences like those in (1) are canceled.

 Nonetheless, the volume Casimir effect (VCE) can play an important
 role in cosmology. In Ref.[4] it is demonstrated that due to VCE
 the matter can transform into a "pseudovacuum" state, which could
 be associated with a source for the Kerr-Newman metric. Moreover in
 that paper it is noted that VCE may play a role in the collapse
 and singularity problems in cosmology.

 In the present work we aim to demonstrate that VCE can lead to
 a huge energy output during the compression of  a sufficiently
 big volume of matter.
 We shall apply it to such astrophysical objects as white dwarfs
 and neutron stars (supernova).
 In what follows, we shall not consider the problem of
 the energy output due to nuclear reactions. As a rule, in our
 calculations, the energy due to VCE exceeds nuclear energy  output.

 Let us consider a volume $V$ that is filled by a real
 conductive material. Then we are waiting to obtain a finite result for
 $\Delta \epsilon_{vac}$. This is a result of the
 well-known fact (e.g., [2,3])
 that real conductors are transparent for high-frequency electromagnetic
 field provided $\omega > \omega_{p}$ , where
 $\omega_p = \sqrt{{{4\pi n e^2} \over {m c^2}}}$ is the plasma frequency,
 $e$ is the charge of electron, $m$ their mass, $n$ the density,
 $c$ is the velocity of light.
 It should be noted that an exact theory for the calculation of VCE for
 a real conductor is absent as yet. In what follows, we shall
 be interested in the estimation of the value of the effect.
 Let us begin with a simple approximation:
 let any good conductive material be an ideal conductor for
 $\omega < \omega_{p}$
 and be a transparent material for $\omega > \omega_{p}$.
 Thence, one has for the  shift  of vacuum energy density (1)
 inside the volume  (due to the presence of the volume)

 \begin{equation}
 \Delta \epsilon_{vac} =
 -\hbar c \int \limits_0^{\omega_{p}} \omega
 \frac{d^3 {\bf k}}{(2\pi)^3} = -{{\omega^4_{p}}\over {4\pi^2}} \hbar c \; ,
 \end{equation}

 Because $\omega_{p}^4 \sim n^2$, the total energy
 ($\Delta \epsilon_{vac}V$) turns out
 to be proportional to $n^2$. During the gravitational
 collapse of a star, the electron density can
 increase dramatically. Therefore one may expect
 some energy release. To find the quantity of this energy,
 let us consider a more accurate calculation. We shall
 address Ref.[5] where there is a rigorous quantum
 calculation of e.m. zero-point energy inside a plane superconductor. Being
 interested essentially in the contribution of the high frequency e.m.
 modes, we can omit the difference in the vacuum solutions
 of e.m. field for the case of plan geometry and when spherical.
 Moreover, the difference between super- and usual conductors
 is vanishing for the high frequency.
 Therefore, the formulae of Ref.[5] may be considered for
 our case.

 In calculations of Ref.[5] the interaction of e.m. field
 with matter is introduced by means of the dispersion of
 dielectric permittivity:

 \begin{equation}
 \varepsilon(\omega)= 1- {{\omega_p^2 }\over \omega^2} \;.
 \label{55}
 \end {equation}

 Taking into account the
 penetration of the field into the material
 as well as the contribution from $\omega > \omega_{p}$,
 one obtains the following value [5]

 \begin{equation}
 \Delta \epsilon_{vac} = {\omega_p^2 \hbar c \over 4\pi^2}
 \int \limits_0^{\Lambda} dq\; q^2 \left\{
 {3\over {q+\sqrt{q^2+\omega_p^2}}} - {1\over 2q}\right\}\;,
 \label{e}
 \end{equation}

 \noindent where $q$ is the magnitude of the Euclidean momentum parallel
 to the material surface, $\Lambda$ is some cut-off parameter.

 After integrating, one finds

 \begin{equation}
 \Delta \epsilon_{vac} = {{\omega_p^2 \hbar c}\over
 {4\pi^2}} \left\{-{\Lambda^2\over 4} - {3\Lambda^4\over 4\omega_p^2} +
 \sqrt{\Lambda^2+\omega_p^2}{{2\Lambda^3+\Lambda \omega_p^2 }
 \over 8\omega_p^2} -\frac38 \omega_p^2
 \ln{{{\Lambda + \sqrt{\Lambda^2+\omega_p^2}}\over \omega_p}}
 \right\} \; .
 \label{enn}
 \end {equation}

 Considering the series for (\ref{enn}) in $\omega_p /\Lambda <<1$, one has

 \begin{equation}
 \Delta \epsilon_{vac} =  {{\omega_p^2
 \Lambda^2 \hbar c}\over {8\pi^2}} -
 {{3\omega_p^4 \hbar c}\over {32\pi^2}}
 \ln{\left({{2\Lambda}\over \omega_p}\right)} +
 {{3\omega_p^4 \hbar c}\over {126\pi^2}} +
 O\left(\omega_p \over \Lambda\right)\;.
 \label{ennn}
 \end {equation}

 The total energy shift arising from
  the presence of conductor (of volume $V$)
  is then

  \begin{equation}
  \Delta E =
  V \Delta \epsilon_{vac} \approx
    {{\hbar N e^2
     }\over {2\pi m c}}\Lambda^2 -
     {{3\hbar N e^2 \omega_p^2}\over {8\pi m c}}
     \ln{\left({{2\Lambda}\over \omega_p}\right)} +
     {{3 \hbar N e^2 \omega_p^2 }\over {32\pi m c}} \; ,
     \label{e4}
     \end {equation}

     \noindent where $N$ is the total number of the electrons
     in the volume $V$.

     The first term, which is proportional to $\Lambda^2$, is
     associated with the electron self-energy. It is precisely that was
     calculated by perturbation theory for the $(e^2/2mc^2)A^2$ term
     (see, e.g. [6]) of the non-relativistic quantum electrodynamics.
     In our calculation to follow, this term will disappear  due to its
     independence of the state of the matter.  The second term
     is of the similar nature as
     the well-known Lamb shift [7]. So, this divergence appears to be
     eliminated by considering the relativistic corrections. To estimate the
     order of magnitude of this term, we shall put $\Lambda \sim 2mc/\hbar$
     as is usual for the Lamb shift calculation. It
     corresponds to the upper limit of where the non-relativistic theory
     is applicable.

     Consider now the following situation. Let the value of
     the electron density $n$ increase
     due to gravitational compression. This leads to the
     following energy creation:

     \begin{eqnarray}
     \delta (\Delta E) &=&\Delta E_{before \ compression}\ \ -\Delta
      E_{after \ compression}
      \nonumber\\ \label{e5}
      &=& {{3\hbar N e^2}\over {8\pi m c}}\left[ f(\omega_p^{after \
compression})
      - f(\omega_p^{before \ compression}) \right],
      \end{eqnarray}

      \noindent where

      $$
      f(\omega)=\omega^2 \left[ \ln{\left({{2\Lambda}\over { \omega}}\right)
      -{1\over 4}} \right] \; .
	$$

	\noindent  Here $\Delta E$ was defined by eq.(\ref{e4}).

	It is important to emphasize here that despite us considering a
	conductive material, what we need to have is the dependence
	(\ref{55}),  which is approximately  valid for any material with
	sufficiently high $\omega$.  So, in principle, this approach
	is valid for non-conductive materials as well.

	For usual (terrestrial) conditions $\delta (\Delta E)$ is a negligible
	quantity. Let us address the problem of compression of a star
	with the solar mass ($M \approx 2\times 10^{30}Kg$). Estimating the
	electron density as $\rho/2m_p$ (helium star), where $\rho$
	is the average density of the star, $m_p$ is the proton mass,
	one has $\omega_p= 2\times 10^{12}m^{-1} (10^6m/R)^{3/2}$.
	To keep $\omega_p < \Lambda \sim 2mc/\hbar \sim 6\times 10^{12}m^{-1}$,
	one needs to consider $R>10^6m$. For such $R$ one can find
	from (\ref{e5}) that

	\begin{equation}
	\delta (\Delta E)  \approx 3 \times 10^{40}J \left({10^6m \over R}\right)^3
	\left\{ \ln{ \left[ 6\left({R\over 10^6m} \right)^{3/2} \right] }
	-{1\over 4} \right \}\;.
	\label{e6}
	\end{equation}

	Recall that the total energy release corresponding to the star explosions
	can reached [8]  $10^{38}J$ for nova, $10^{44}J$ for the supernova
	of type I  and $10^{42}J$ for the supernova of type II.
	If we compare  these values with the energy output (\ref{e6}),
	one finds that the explosion of nova could be explained
	by the compression of a sun-like star to a radius $R\sim 10^7m$.
	It is interesting to note that this radius is about that for
	a white dwarf.

	If we consider the further compression to a neutron star
	(about the latter - see, e.g., [9,10]), we
	will not be able to use the formulae (\ref{enn})-(\ref{e6})
	 because the relativistic
	 corrections (due to the creation of electron-positron pairs by the
	 virtual photons) could be considerable. At least, we do not know
	 which reasonable cut-off parameter $\Lambda$ we should take.
	 The calculation of the relativistic corrections is beyond the scope
	 of this work. Nonetheless, let us note that
	 the relativistic correction to the photon propagator due to
	 one ${\rm e}^+{\rm e}^-$ loop is about [11]
	 $2\alpha/(3\pi)\times \ln{(q/m)}$. It is a small value for
	 practically any reasonable $q$. Therefore, one can hope
	 that the relativistic correction will be small.
	 So, let us estimate the possible energy release due to
	 the further compression to a neutron star by the help of eq.(\ref{e5}).
	  If we reach a
	  radius $R\sim 6.5\times 10^4m$, the energy output
	  can be equal to  $\delta (\Delta E)  \approx 10^{44}J $
	  (here we put $\Lambda > 40mc/\hbar$).

	  To reach the scales of neutron stars, it is necessary
	  to compress the star, at least, to $3\times 10^4m$.
	  From the formula (\ref{e5}), it should lead to some increasing
	  the energy output. On the other hand, the quantity of
	  electrons (N) would decrease due to transformation
	  (with protons) to neutrons. Moreover the relativistic
	  corrections could be considerable.

	  So, the energy obtained above should be considered as the
	  estimation of the energy output, at least, during the
	  creation of neutron star. Comparing the energy release
	  during the supernova explosion (see above) with the
	  estimated value, one can
	  conclude that the mechanism suggested could be
	  the explanation of these explosions.

	  Consider now quasars or galaxy cores.
	   It is reasonable to assume that some
	   process of collapse occurs in the core of the quasar.
	   Let a neutron-star-object
	   be the result of such a collapse. As was demonstrated above,
	    it leads
	    to  powerful energy output. In contrast to the previous case, there will
be
	    a strong "gravitational wall" around the neutron-star-object.
	    If these two factors compensate each other, the processes will be
	    in equilibrium.
	    The process of adding new material to the neutron-star-object
	    can continue as soon as a temperature balance is reached between the
	    neutron-star-object and the rest of the quasar. The excess of
	     material of the
	     neutron-star-object could collapse to a black hole,
	      adding more gravitational
	      energy to quasar (the question about the
	      more compression than neutron star is unclear for now).

	      Estimate now, how long a quasar can
	      shine due to the process described above.
	      Let us consider the neutron-star-object
	      using the same method we employed for
	      the neutron star. We find that  one half of the quasar
	      matter could produce
	      a quasar radiation output of $\sim 10^{41}J/s$ (maximal
	      observable quasar power [8,12]) for the time

	      \begin{equation}
	      t=1.3\times 10^5 yr\;,
	      \end{equation}

	      The mass
	      of the quasar is taken to be $10^9$ solar masses.

	      All these estimates show that the volume Casimir effect
	      should be taken into account in Astrophysics and Cosmology.
	      I hope that this notice encourages the efforts
	      to elaborate of the theory of the volume Casimir effect.

	      I would like to thank A.Ya.Burinskii, G.Starkman and
	      J.W.Moffat  for useful discussions.
	      Moreover, I am grateful to J.W.Moffat for support
	      during the complition this work and N.J.Cornish for carefully
	      reading the manuscript.

	      \newpage

	      \centerline{References}

	      1.H.B.G.Casimir, Physica XIX, 846 (1956).

	      2.N.D.Birell and P.C.W.Davies, {\it Quantum fields in curved spacetime},

	      (Cambridge U.P., Cambridge ,1982).

	      3.N.N.Trunov and V.M.Mostepanenko, {\it Casimir effect and its
applications},

	      (Cambridge U.P., Cambridge, 1993).

	      4.A.Ya.Burinskii, Phys.Lett.B216, 123 (1989).

	      5.T.Kunimasa, Prog.Theor.Phys. 83, 97 (1990).

	      6.E.A. Power, {\it Introductory Quantum Electrodynamics},
	      (Longmans Green

	      and Co.,Ltd., London, 1964).

	      7. T.A.Welton, Phys.Rev. 74, 1157 (1948).

	      8.C.W.Allen, {\it Astrophysical quantities}, (The Athlone Press, London,
1973).

	      9.G.Baym and C.J.Pethick, Ann.Rev.Nucl.Sci. 25, 27, (1975);

	      Ann.Rev.Astron.a.
	      Astrophys. 17, 415 (1979).

	      10.F.K.Lamb, Ann.New York Acad.Sci. 302, 482 (1977).

	      11.C.Itsikson and J.-B.Zuber, {\it Quantum Field Theory},

	      (McGraw-Hill Book Company, Ney York, 1980).

	      12.V.L.Ginzburg, {\it On physics and astrophysics}, (Nauka, Moscow,
1985). {\it
	      In Russian}.

	      \end{document}